\documentclass[a4paper]{article}
\usepackage{graphicx,amssymb,color}
\usepackage[colorlinks=true, urlcolor=blue]{hyperref}
\usepackage{epstopdf}

\begin{document}

\title{Complex transitions to synchronization in delay-coupled networks of logistic maps}

\author{Cristina Masoller\thanks{Departament de F\'{\i}sica i Enginyeria Nuclear, Universitat Polit\`{e}cnica de Catalunya, Colom 11, Terrassa 08222, Barcelona, Spain. \texttt{cristina.masoller@gmail.com}} \and Fatihcan M. Atay\thanks{Max Planck Institute for Mathematics in the Sciences, 04103 Leipzig, Germany. \texttt{fatay@mis.mpg.de}}}

\date{\emph{Preprint.} Final version in \emph{European Physical Journal}~D 62:119--126, 2011. \smallskip
\hspace*{25mm} \href{http://dx.doi.org/10.1140/epjd/e2011-10370-7}{DOI: 10.1140/epjd/e2011-10370-7}\hspace*{\fill}
}

\maketitle

\abstract{A network of delay-coupled logistic maps exhibits two different synchronization regimes, depending on the distribution of the coupling delay times. When the delays are homogeneous throughout the network, the network synchronizes to a time-dependent state [Atay et al., Phys. Rev. Lett. {\bf 92}, 144101 (2004)], which may be periodic or chaotic depending on the delay; when the delays are sufficiently heterogeneous, the synchronization proceeds to a steady-state, which is unstable for the uncoupled map [Masoller and Marti, Phys. Rev. Lett. {\bf 94}, 134102 (2005)]. Here we characterize the transition from time-dependent to steady-state synchronization as the width of the delay distribution increases. We also compare the two transitions to synchronization as the coupling strength increases. We use transition probabilities calculated via symbolic analysis and ordinal patterns.  We find that, as the coupling strength increases, before the onset of steady-state synchronization the network splits into two clusters which are in anti-phase relation with each other. On the other hand, with increasing delay heterogeneity, no cluster formation is seen at the onset of steady-state synchronization; however, a rather complex unsynchronized state is detected, revealed by a diversity of transition probabilities in the network nodes.
}


\maketitle

\section{Introduction}
A fascinating and intriguing feature of spatially extended systems composed of many interacting units, like chanting crowds, tropical Malaysian flashing fireflies, pacemaker heart cells, cells governing the circadian rhythms, pedestrians crossing the Millennium Bridge, etc., is that they can synchronize even when the units are spread over wide spatial areas \cite{glass,millenium_bridge,sincro_comp_net}. In order to understand their synchronization phenomena, these systems have been modeled by networks of coupled phase oscillators, like the Kuramoto model \cite{kuramoto}, and by networks of coupled maps \cite{kaneko,jalan_2003,kurths_pre_2009}, such as circle maps \cite{batista_2003,circle_maps}, Bernoulli maps \cite{kinzel_pre_2009}, logistic maps \cite{gallas_2006,kanter_pre_2009,ponce_2009}, Rulkov maps \cite{chen_2008,chen_2009,kurths_ijbc_2010} etc.

In systems of coupled units, communication delays naturally arise from a realistic consideration of the finite speed of information transmission between pairs of units, and can have a great impact on their collective behavior. In particular, in networks of coupled maps, synchronization phenomena in the presence of time-delays has received considerable attention and is still an active research area \cite{kinzel_pre_2009,gallas_2006,kanter_pre_2009,ponce_2009,chen_2008,chen_2009,atay_prl03,atay_2010}. Networks of delayed-coupled maps are popular for studying the effects of delayed interactions because one can simulate large ensembles of coupled units, even in the presence of heterogeneous and long delays, with a great reduction of computational time and memory requirements, as compared to delay-differential rate-equations. The logistic map has been a popular choice because is a prototype example of how chaotic dynamics and universal scaling laws \cite{f1,CT,TC,f2} arise in simple non-linear systems.

In networks of delayed-coupled logistic maps, when the delays are heterogeneous the network exhibits a synchronized collective behavior that is qualitatively different from that of instantaneously interacting units, or by units interacting with homogeneous delays \cite{atay_PRL_2004,prl}. Heterogeneous delays can enhance the synchronizability of the network, but they can also affect its synchronized dynamics. A network of delayed coupled logistic maps displays two qualitatively different synchronization regimes, depending on the delay distribution. When the delays are homogeneous throughout the network, the network synchronizes to a time-varying state \cite{atay_PRL_2004}, and the synchronizability depends mainly on the network architecture; when the delays are sufficiently heterogeneous, the network synchronizes to a steady-state, which is unstable for the uncoupled maps \cite{prl}, and the synchronizability depends mainly on the average number of neighbors per node.

The stability of the steady-state of delay-coupled maps is well-understood when the delay is homogeneous (delta-distributed): Ref.~\cite{atay_siads06} gave exact conditions for stability and showed that the largest eigenvalue of the Laplacian matrix determines the effect of the network structure on stability. Such precise results are unavailable for arbitrary delay distributions. Nevertheless, it is known that distributed delays can induce or improve stability of the steady-state in coupled limit-cycle oscillators \cite{atay_prl03}, or in more general delay-differential equations in the vicinity of a Hopf instability \cite{atay_dcds08}. A recent example is reported in \cite{omi_pre_2008}, for an integro-differential equation describing the collective dynamics of a neural network with distributed signal delays: With Gamma distributed delays, which are less dispersed than the exponential distribution, the system exhibits reentrant phenomena (i.e., the stability is lost but then recovered as the mean delay is increased), while with delays that are more highly dispersed than exponential, the system does not destabilize.

The aim of this paper is to characterize the transition to the two synchronized regimes of delayed coupled logistic maps (time-dependent for homogeneous delays and steady-state for heterogeneous ones) as the coupling strength or as the width of the delay distribution increases. The degree of synchronization is measured in terms of the transition probabilities in the network nodes, which are calculated via symbolic analysis and ordinal patterns. The symbolic method is based in dividing the state space of a given node into two regions and considering the relative frequencies of the transitions between those regions \cite{atay_chaos_2006}; the ordinal patterns method is based in defining patterns in the time-series of a node that result from ordering relations in consecutive values in the series \cite{bp}, and computing the relative frequencies of the transitions between those patterns. The paper is organized as follows: Section II presents the network model and the magnitudes employed to quantify the degree of synchronization. Section III presents the results, and Sec. IV contains a summary and the conclusions.

\section{The model and synchronization quantifiers}

We consider $N$ logistic maps coupled as
\begin{equation}
\label{array}
 x_i(t+1)= f[x_i(t)] + \frac {\epsilon} {k_i}\sum_{j=1}^N
w_{ij}\left(f[x_j(t-\tau_{ij})] - f[x_i(t)]\right),
\end{equation}
where $t$ is a discrete time index, $i$ is a discrete spatial index, $f(x)=ax(1-x)$ is the logistic map with parameter $a$, $\epsilon$ is the coupling strength,
 $\tau_{ij}$ denotes the delay in the link from node $j$ to $i$, $w_{ij}$ are the elements of the adjacency matrix $w$ whose values equals 1 if there is a link from node $j$ to node $i$ and 0 otherwise, and $k_i$ is the in-degree of the node $i$, $k_i =\sum_j w_{ij}$. Here, $\tau$ and $w$ are not restricted to be symmetric matrices.

When the delays are sufficiently heterogeneous, the solution in the spatially homogeneous steady-state,
\begin{equation}
\label{steady-state}
x_i(t)=x_0 \texttt{  } \forall i,
\end{equation}
is stable in a certain range of coupling strengths \cite{prl}, where $x_0$ is the fixed point of the uncoupled logistic map,
\begin{equation}
\label{fixed_p}
x_0=f(x_0)=1-1/a.
\end{equation}
We will refer to this solution as ``steady-state synchronization". In contrast, when the delays are homogeneous throughout the network  ($\tau_{ij}=\tau_0$ $\forall i,j$) the network synchronizes to a time-dependent state \cite{atay_PRL_2004},
\begin{equation}
\label{time-dependent}
x_i(t)=x(t) \texttt{  } \forall i,
\end{equation}
where $x(t)$ is a solution of
\begin{equation}
 x(t+1)= f[x(t)] + \epsilon \left(f[x(t-\tau_{0})] - f[x(t)]\right),
\end{equation}
and thus, the dynamics can be periodic or chaotic depending on $\tau_0$. We will refer to this situation as ``time-dependent" synchronization.

Clearly, other ``out of phase" synchronization regimes, where the different nodes maintain certain lag-times among them, are also possible. For example, a 1D linear globally-coupled network with distance-dependent delays, $\tau_{ij}=|i-j|/v$, where $v$ is the speed of information transmission, synchronizes to a state in which the nodes evolve along a periodic orbit of the uncoupled logistic map (i.e., $x_i(t)$ is a solution of $x_i(t+1)=f[x_i(t)]$), while the spatial correlation of the nodes along the network is such that $x_i(t) = x_j(t-\tau_{ij})$ $\forall i,j$ (i.e., each map ``sees" all other maps in his present, current, state) \cite{marti,marti2}.
In the following we only focus on ``steady-state" and ``time-dependent" synchronization.

To capture the degree of synchronization and to distinguish between steady-state synchronization, Eq. (\ref{steady-state}), and time-dependent synchronization, Eq. (\ref{time-dependent}), we use the following measures:

1) The variance of the nodes' states,
\begin{eqnarray}
\label{sp1}
 \sigma^2 &=& \frac{1}{N} \langle \sum_{i=1}^{N} \left(x_i(t)- \langle x \rangle_s\right)^2
 \rangle_t
\end{eqnarray}
where $\langle.\rangle_s$ denotes an average over the nodes of the network, and $\langle . \rangle_t$ denotes an average over time.

2) The variance of the distance to the steady state,
\begin{eqnarray}
\label{sp2}
 \sigma'^2 &=& \frac{1}{N} \langle \sum_{i=1}^{N} \left(x_i(t)- x_0\right)^2 \rangle_t,
\end{eqnarray}
where $x_0$ is the fixed point of the uncoupled logistic map, Eq. (\ref{fixed_p}).

One can notice that $\sigma^2 =0$ if and only if $x_i=x_j=\langle x \rangle_s$ $\forall i,j$, while $\sigma'^2 =0$ if and only if $x_i=x_0$ $\forall i$. Thus, $\sigma'^2$ allows to distinguish synchronization in the steady state from synchronization in a time dependent state. In the former case, both $\sigma^2$ and $\sigma'^{2}$ are zero, in the latter case, only $\sigma^{2}$ is zero.

We note that both $\sigma^2$ and $\sigma'^2$ are ``global" indicators that give no information about the microscopic local dynamics in the nodes of the network.
To gain inside into this local dynamics, the transition probabilities in individual nodes can be computed via symbolic dynamics \cite{atay_chaos_2006} or ordinal patterns \cite{bp}, as follows.

3) Transition probabilities computed via symbolic dynamics: At each node $i$, a two-symbol dynamics is generated by the partition of the phase space as
\begin{eqnarray}
s_i(t)&=&\alpha \, \mbox{ if } x_i(t) \leq x^* \nonumber \\
s_i(t)&=&\beta  \, \mbox{ otherwise, }
\end{eqnarray}
where $x^*$ is a threshold value, which in the following is chosen equal to the fixed point of the uncoupled logistic map, $x_0$.
The transition probability in node $i$, $P_{i,sd}(\alpha,\alpha)$, is calculated as
\begin{equation}
P_{i,sd}(\alpha,\alpha)= \frac{\sum_{t=1}^L n(s_i(t)=\alpha, s_i(t+1)=\alpha)}{\sum_{t=1}^L n(s_i(t)=\alpha)},
\end{equation}
where $n$ is a count of the number of times of occurrence in a time-series of length $L$.
The global properties of the network can be quantified by the variance of $P_{i,sd}(\alpha,\alpha)$ over the network \cite{atay_chaos_2006},
\begin{equation}
 \zeta^2_{sd} = \frac{1}{N} \sum_{i=1}^{N} (P_{i,sd}(\alpha,\alpha)- \langle P_{sd}(\alpha,\alpha)\rangle_s)^2 ,
\end{equation}
where $\langle P_{sd}(\alpha,\alpha)\rangle_s=(1/N)\sum_{i=1}^{N} P_{i,sd}(\alpha,\alpha)$ is the average transition probability.

4) In addition, in each node $i$, a sequence of symbols can be generated via a comparison of consecutive values (``ordinal patterns" of dimension two, as proposed by Brandt and Pompe \cite{bp})
\begin{eqnarray}
s_i(t)&=&\alpha \, \mbox{ if } x_i(t) \leq x_i(t+1) \nonumber \\
s_i(t)&=&\beta  \, \mbox{ otherwise. }
\end{eqnarray}
A nice advantage of this procedure is that it does not require the definition of a threshold. As before, the transition probability in node $i$, $P_{i,BP}(\alpha,\alpha)$, can be calculated as
\begin{equation}
P_{i,BP}(\alpha,\alpha)= \frac{\sum_{t=1}^L n(s_i(t)=\alpha, s_i(t+1)=\alpha)}{\sum_{t=1}^L n(s_i(t)=\alpha)},
\end{equation}
and its variance,
\begin{equation}
 \zeta^2_{BP} = \frac{1}{N} \sum_{i=1}^{N} (P_{i,BP}(\alpha,\alpha)- \langle P_{BP}(\alpha,\alpha)\rangle_s)^2,
\end{equation}
where $\langle P_{BP}(\alpha,\alpha)\rangle_s=(1/N)\sum_{i=1}^{N} P_{i,BP}(\alpha,\alpha)$, can be used to capture global properties of the network.

\graphicspath{{c:/fortran/mapas/}}
\begin{figure}[tbp]
\begin{center}
\resizebox{1.0\columnwidth}{!}{\includegraphics {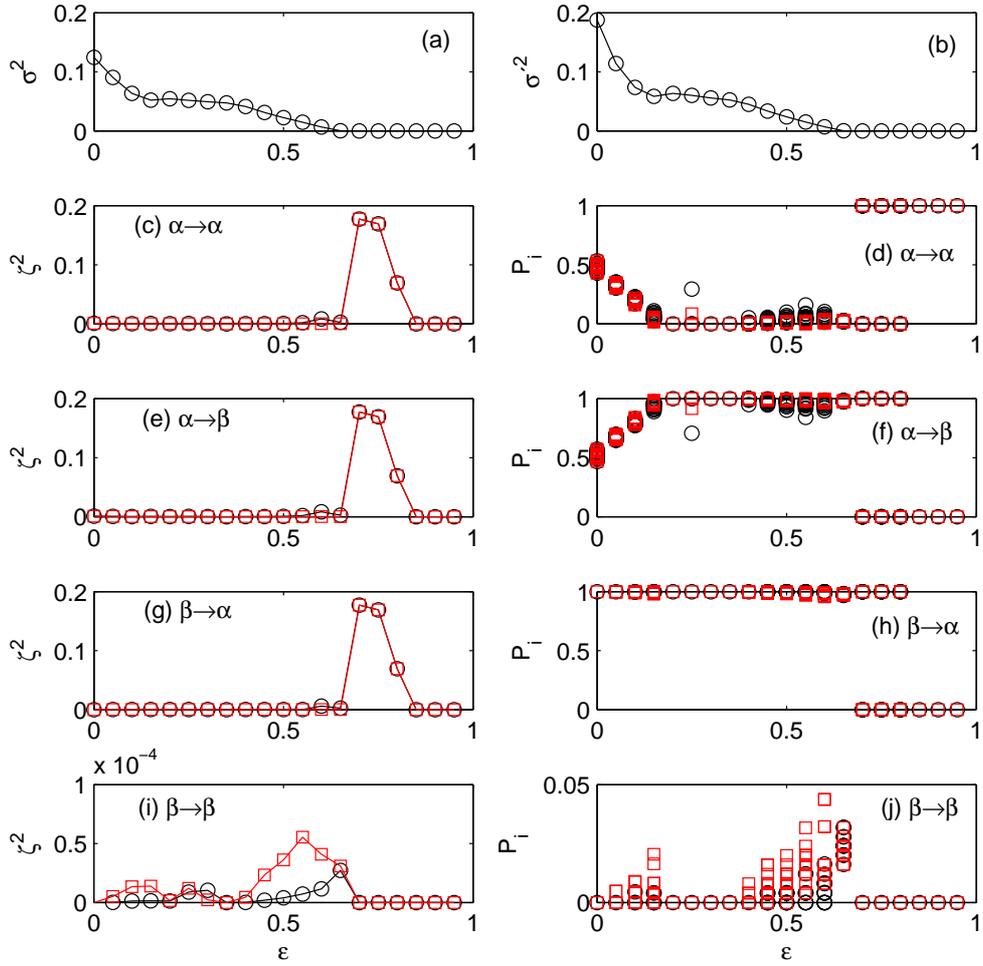}}
\caption{Transition to ``steady-state" synchronization for fixed delay heterogeneity ($c^*=0.6$) and increasing coupling strength, $\epsilon$. The quantifiers $\sigma^2$, $\sigma'^2$, and $\zeta^2$ are plotted vs. the coupling strength. The transition probabilities $P_i$ in 20 randomly selected nodes are also shown. In panels (c)-(j) the transition probabilities are computed via symbolic dynamics (circles) and ordinal patterns (squares; red online).}
\label{fig:1}
\end{center}
\end{figure}

\section{Results}

In the following we present the results for an Erd{\"o}s-Renyi random network \cite{random} of $N$ nodes with an average degree $\langle k \rangle_s$ such that the network has a single component. Unless otherwise explicitly stated, $N=200$, $\langle k \rangle_s=20$ and the coupling delays are Gaussian distributed with a mean delay $\langle \tau \rangle_s=5$. The parameter that controls the delay heterogeneity is the standard deviation of the delay distribution, normalized by the mean delay, $c^* = \sigma_\tau/\langle \tau \rangle_s$. The parameter of the logistic map is taken to be $a=4$ and the simulations start with random initial conditions. Unless otherwise explicitly stated, the quantifiers $\sigma^2$, $\sigma'^2$, $\zeta^2_{sd}$ and $\zeta^2_{BP}$ are computed over time series of length $L=500$, after the first $3000$ iterations are disregarded, and they are averaged over 20 stochastic trajectories, where the random initial conditions ($x_i(0)$), delay distribution ($\tau_{ij}$), and adjacency matrix ($w_{ij}$) are varied.

First we consider the transition to ``steady-state" synchronization as the coupling strength $\epsilon$ increases, while the delay heterogeneity $c^*$ is kept constant. The delays are sufficiently heterogeneous such that, for large enough $\epsilon$, the network synchronizes as $x_i=x_0$ $\forall i$.

Figure 1 displays $\sigma^2$, $\sigma'^2$, $\zeta^2_{sd}$ and $\zeta^2_{BP}$ vs. the coupling strength $\epsilon$. It also displays the four transition probabilities, for one typical stochastic trajectory, in 20 randomly selected nodes, as computed via symbolic dynamics (circles) and ordinal patterns (squares). It can be seen that before the onset of synchronization there is a formation of two clusters, as the transition probabilities $P_{i}(\alpha,\alpha)$, $P_{i}(\alpha,\beta)$ and $P_{i}(\beta,\alpha)$ are 0 in some nodes and 1 in others. One can also notice that $P_{i}(\beta,\beta)$ is very small in all the nodes, and that the transition probabilities calculated with symbolic dynamics are very similar to those calculated with ordinal patterns.

Further insight into the networks' dynamics near the synchronization transition can obtained by examining the time evolution of the quantifiers, of the transition probabilities (now computed over a moving time-window of length 500), and the dynamics of a few, randomly selected nodes. These are shown in Fig.~2, where the coupling strength is slightly smaller than that needed for "steady-state" synchronization. In Fig.~2(d) the network configuration at a fixed time (i.e., a 'snapshot' of the states of the nodes) is also shown. One can notice that the nodes form two clusters that oscillate in anti-phase: when one cluster is above the fixed-point solution, the other one is below, and at the next time step, the two clusters switch their positions.

Next, we consider the situation where the delay heterogeneity $c^*$ increases, starting with a delay distribution that is a delta function ($c^*=0$), while the coupling strength $\epsilon$ is kept constant. The coupling strength is strong enough that, for homogeneous delays, the network synchronizes as $x_i=x_j$ $\forall i,j$ (time-dependent synchronization), while for sufficiently heterogeneous delays, the network synchronizes as $x_i=x_0$ $\forall i$ (steady-state synchronization). Figure 3 displays the quantifiers vs. the delay heterogeneity and also displays the four transition probabilities for one typical stochastic trajectory, in 20 randomly selected nodes. In this scenario, for small delay heterogeneity the time-dependent synchronization is gradually lost, and as the delay heterogeneity increases, there is a smooth transition to the steady-state synchronization. No cluster formation can be observed at the onset of ``steady-state synchronization", since the transition probabilities are within a certain range of values.

The dynamics of the network near ``steady-state" synchronization is examined in Fig.~4, with parameters such that the heterogeneity of the delays is slightly smaller than that needed for ``steady-state" synchronization. Here one can notice that the nodes evolve together, in a single cluster, displaying slow oscillations around the steady state [compare the oscillation frequencies in Figs. 2(f),(h),(j) with 4(f),(h),(j)]. The period and shape of these oscillations vary with $c^*$. One should keep in mind that the scenario we are considering is with strong coupling, such that, for $c^*=0$ the array synchronizes in a time-dependent state; the network dynamics near this state (with the presence of a small delay heterogeneity), is shown in Fig. 5.

The approach towards ``steady-state synchronization" reveals `critical slowing down' in the sense that the amplitude of the oscillations in Figs.  2(f),(h),(j) and 4(f),(h),(j) gradually decreases with increasing $\epsilon$ or $c^*$, and there is a slow approach towards the fixed-point solution. The main differences being that for sufficiently heterogeneous delays and small coupling, the network splits in two clusters which display fast anti-phase oscillations, while for large enough coupling but not sufficiently heterogeneous delays, the network approaches the fixed point solution as a single cluster and slow oscillations.

\begin{figure}[tbp]
\begin{center}
\resizebox{1.0\columnwidth}{!}{\includegraphics {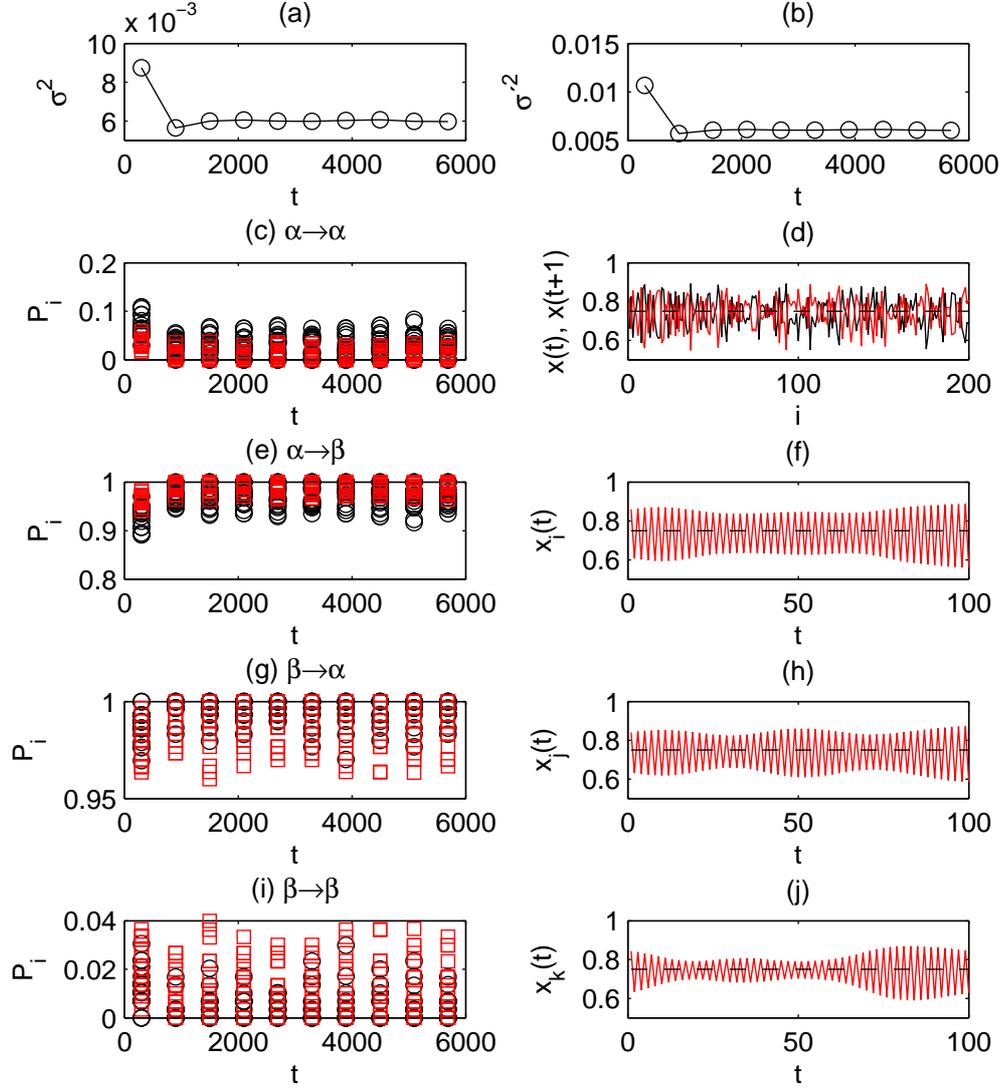}}
\caption{Time-variation of the quantifiers  $\sigma^2$ (a), $\sigma'^2$ (b), and of the transition probabilities (c), (e), (g), (i), calculated in a moving time-windows of length L=600. The circles indicate transition probabilities computed via symbolic dynamics; the squares (red online), via ordinal patterns. The parameters are such that the coupling strength is slightly below that required for synchronization in the steady-state ($\epsilon=0.6$ and $c^*=0.6$). Panel (d) displays the network configuration at two consecutive times, and panels (f), (h), (j), the time evolution of three randomly selected nodes. In (d),(f),(h) and (j) the dashed horizontal line indicates the fixed point $x_0$ of the map $f$.}
\label{fig:2}
\end{center}
\end{figure}

\begin{figure}[tbp]
\begin{center}
\resizebox{1.0\columnwidth}{!}{\includegraphics {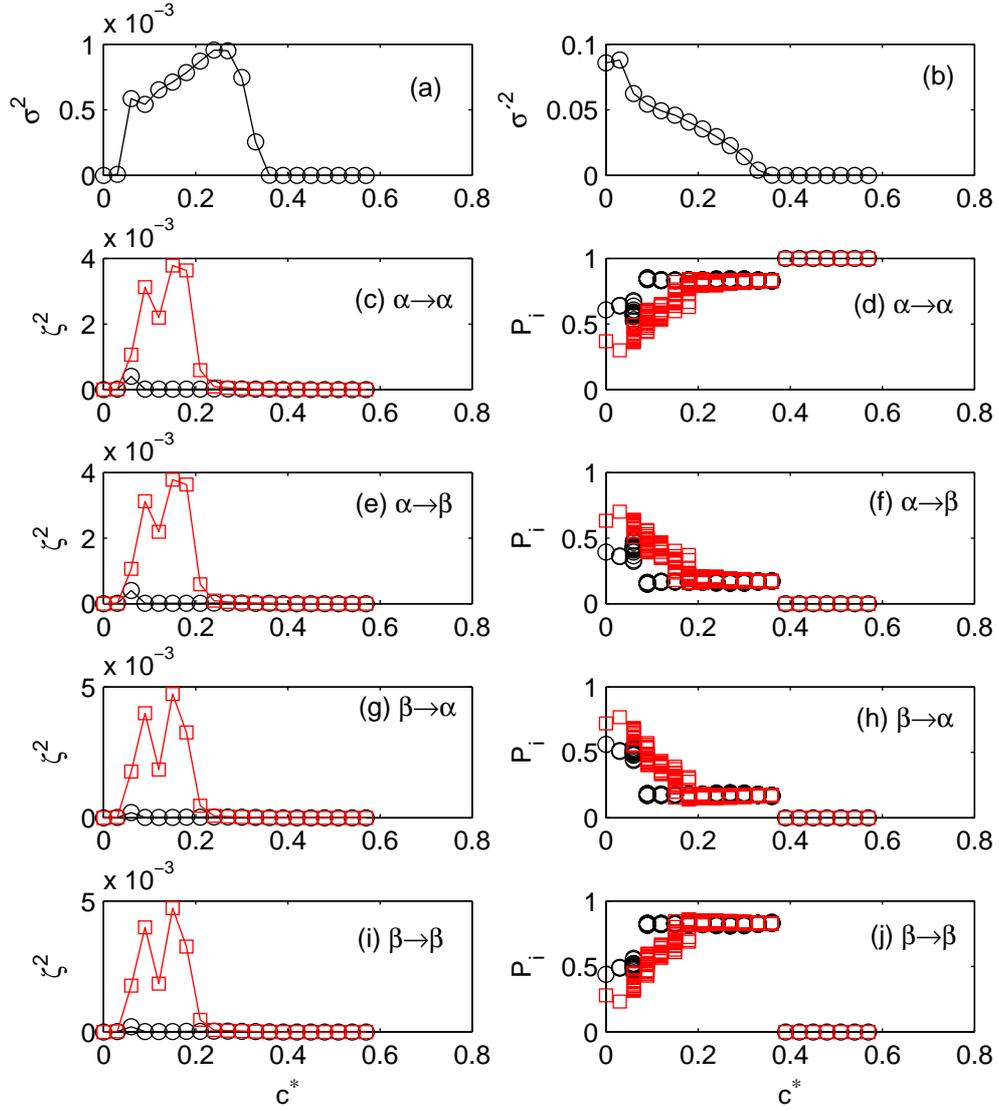}}
\caption{As Fig.~1, but keeping the coupling strength fixed ($\epsilon=0.9$) and increasing the delay heterogeneity $c^*$.}
\label{fig:3}
\end{center}
\end{figure}

\begin{figure}[tbp]
\begin{center}
\resizebox{1.0\columnwidth}{!}{\includegraphics {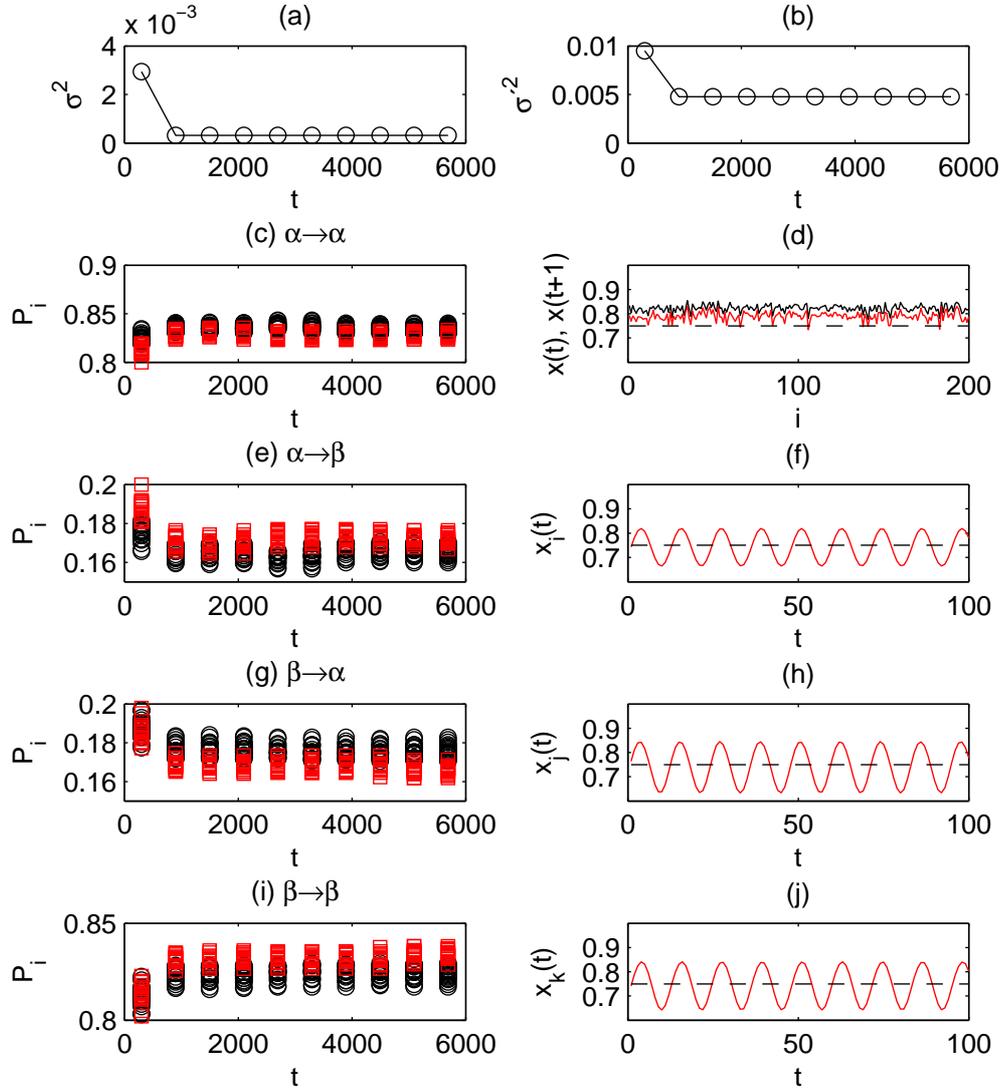}}
\caption{As Fig.~2 but with the delay heterogeneity slightly below that required for synchronization to the steady-state ($\epsilon=0.9$, $c^*=0.33$).}
\label{fig:4}
\end{center}
\end{figure}

\begin{figure}[tbp]
\begin{center}
\resizebox{1.0\columnwidth}{!}{\includegraphics {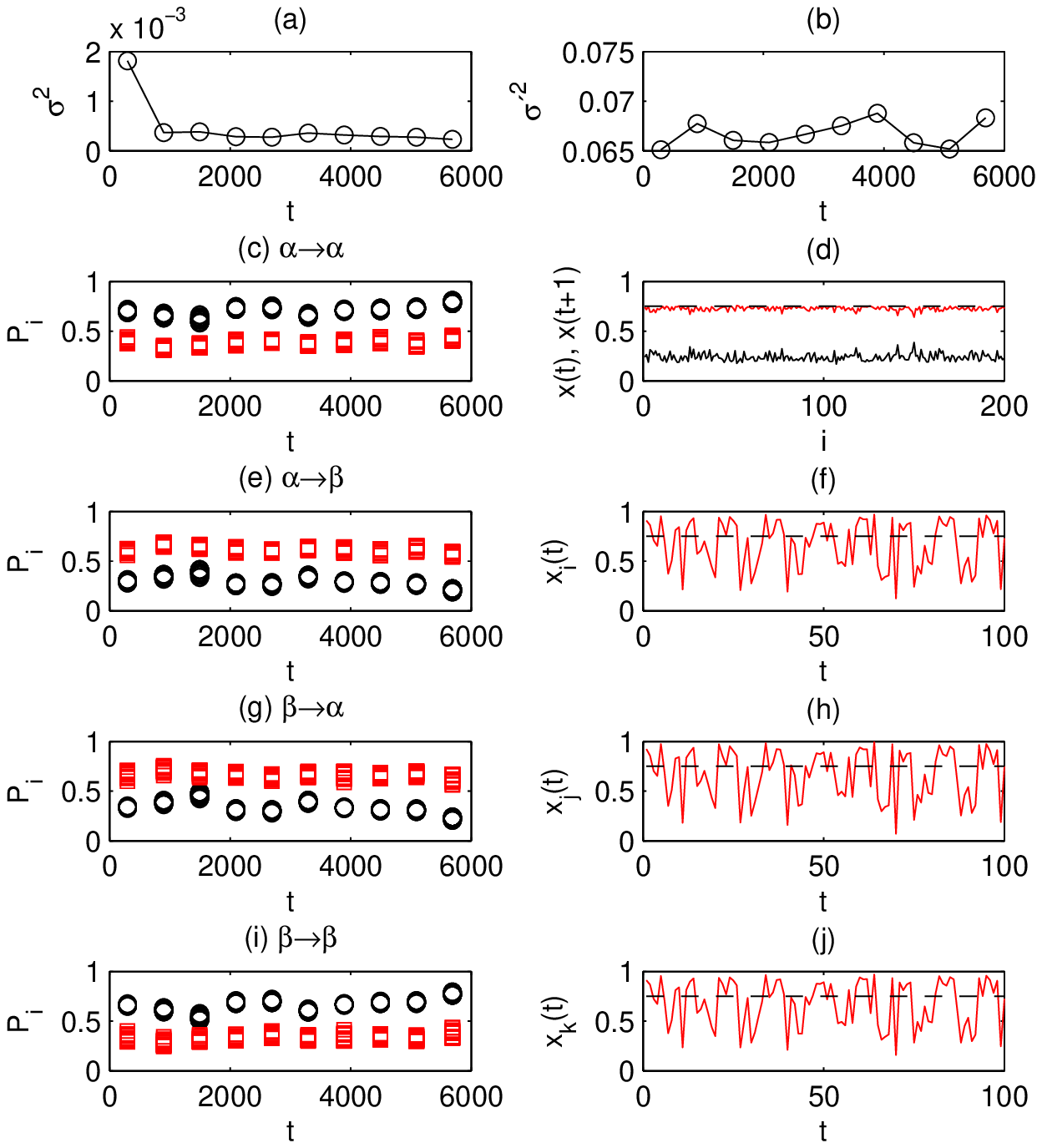}}
\caption{As Fig.~4 but for small delay heterogeneity ($\epsilon=0.9$, $c^*=0.05$).}
\label{fig:4bis}
\end{center}
\end{figure}

One can then interpret the diversity of transition probabilities seen at the boundary of steady-state synchronization as ``noise amplification". When the network is almost or nearly synchronized, for all the nodes we have $x_i(t)\sim x_0$ and therefore very small variations near $x_0$ result in a diversity of transition probabilities. This occurs when both $\epsilon$ or $c^*$ is varied.
However, because of the different way the network approaches the homogeneous solution, increasing $\epsilon$ yields two clusters and the transition probabilities are either close to 0 or to 1, while, increasing $c^*$ yields a single cluster and the transition probabilities are within an interval of values.

\begin{figure}[tbp]
\begin{center}
\begin{tabular}{c}
\includegraphics[height=4cm, width=\columnwidth]{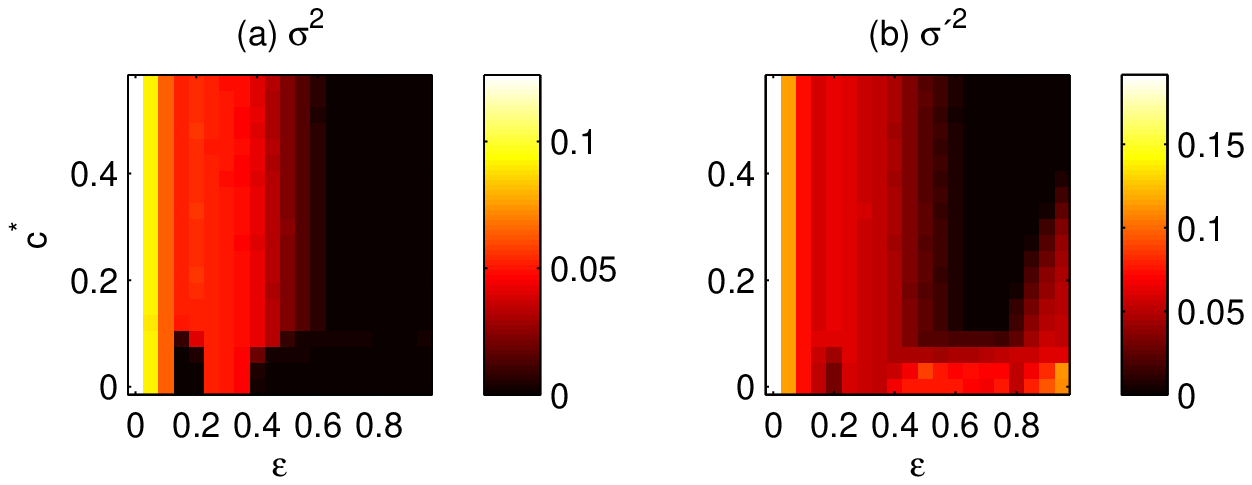}\\
\includegraphics[height=9cm, width=\columnwidth]{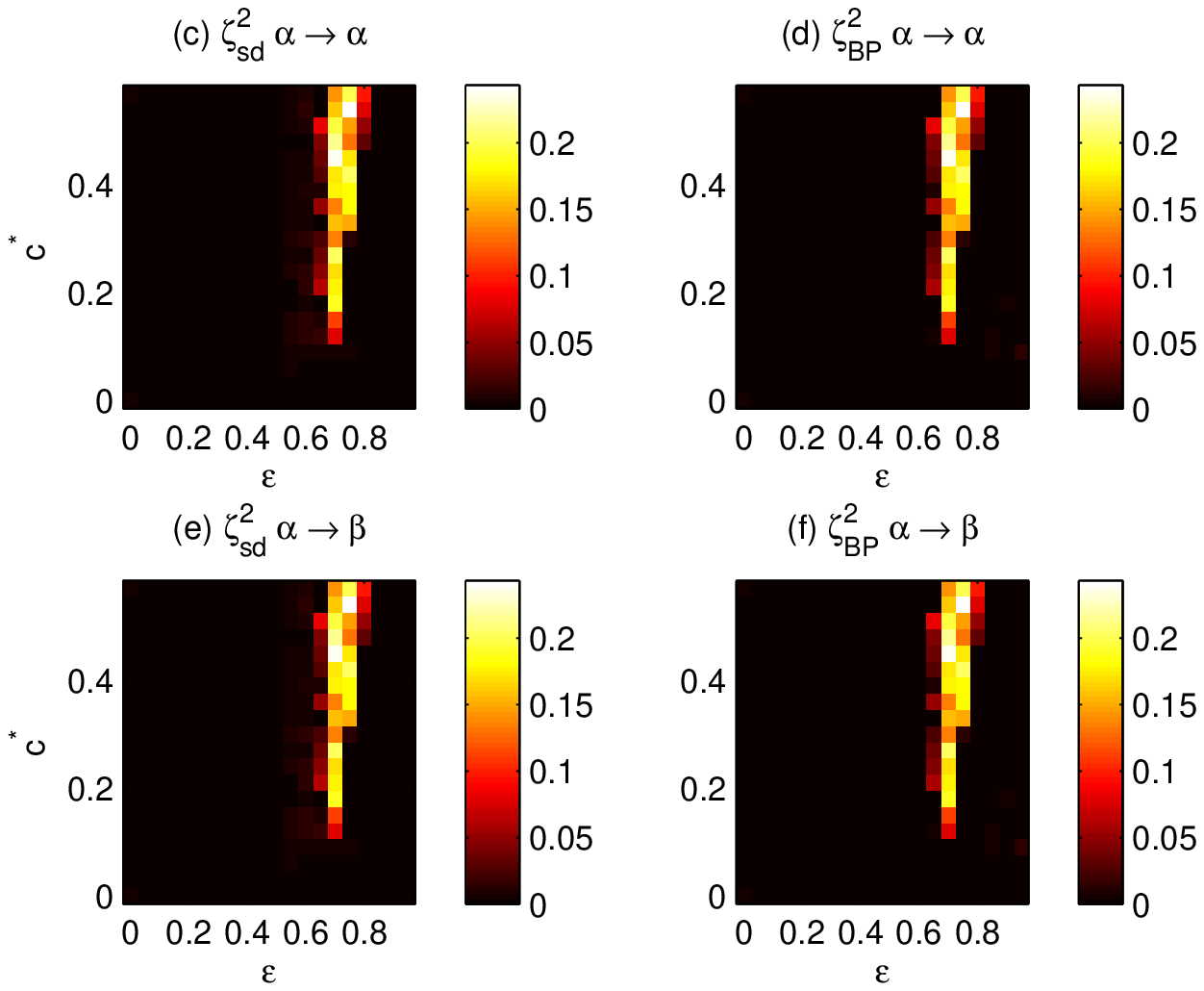}\\
\includegraphics[height=9cm, width=\columnwidth]{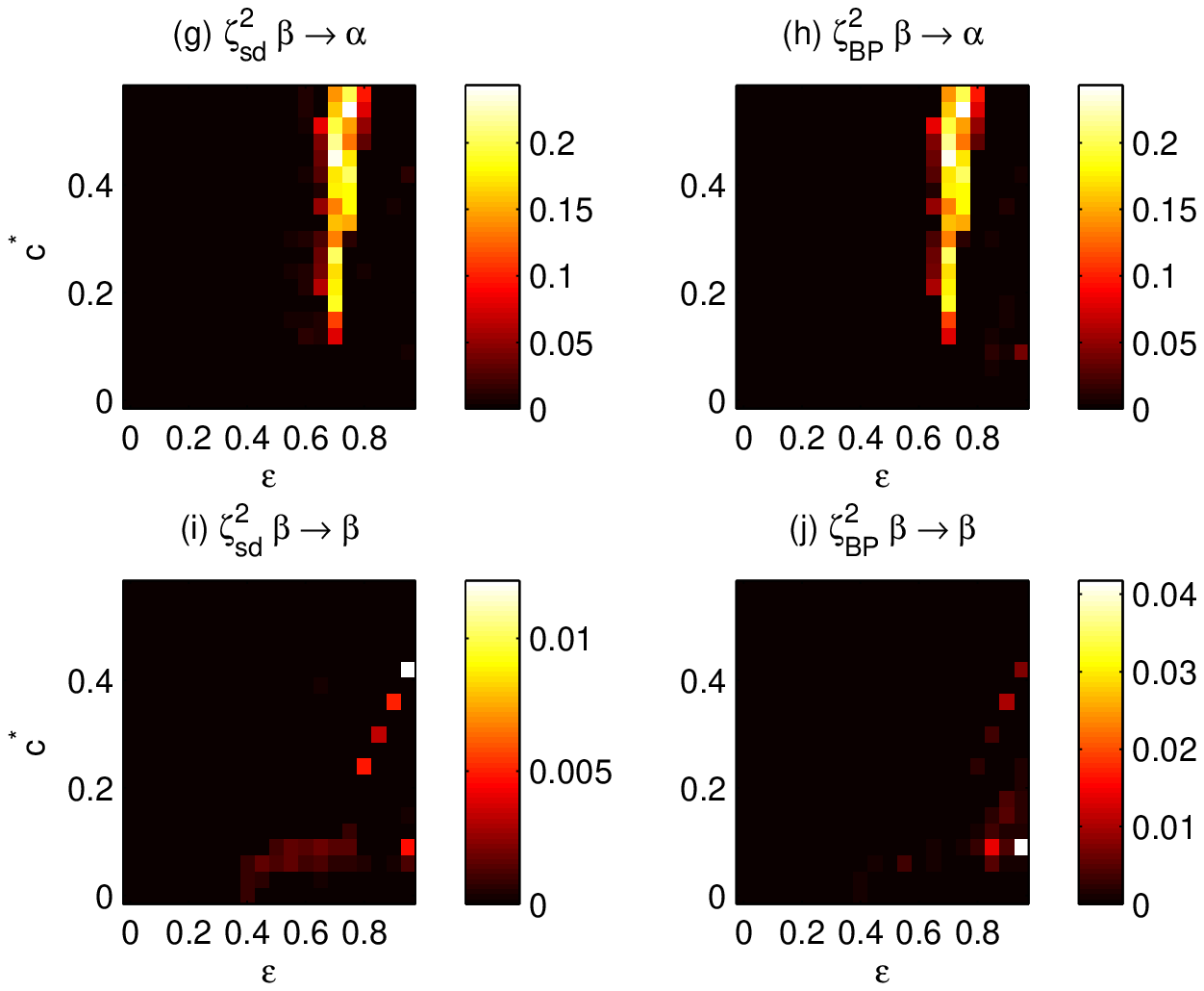}
\end{tabular}
\caption{Synchronization quantifiers in the parameter space (coupling strength on the horizontal axis, delay heterogeneity on the vertical).}
\label{fig:diag_phase}
\end{center}
\end{figure}

Two-dimensional plots in the parameter space (coupling strength, delay heterogeneity), shown in Fig.~\ref{fig:diag_phase}, provide a more complete picture of the various dynamical regimes. We can recognize two synchronization regions occurring for large coupling: steady-state synchronization for large delay heterogeneity [top-right corner in Figs.~\ref{fig:diag_phase}(a),(b), where both $\sigma^2$ and $\sigma'^2$ are zero], and time-dependent synchronization, for homogeneous delays [bottom-right corner in Figs.~\ref{fig:diag_phase}(a),(b), where only $\sigma^2$ is zero].
In addition, there is a narrow window of synchronization for weak coupling strength and almost homogeneous delays [$\eta\sim 0.15-0.2$, bottom-left corner in Figs.~\ref{fig:diag_phase}(a),(b), where $\sigma^2$ is zero and $\sigma'^2$ is small]. This region was reported in \cite{atay_PRL_2004} for homogeneous and odd delay values, and  it can be seen from Fig.~\ref{fig:diag_phase} that it is also robust to small delay heterogeneities.

\begin{figure}[tbp]
\begin{center}
\resizebox{1.0\columnwidth}{!}{\includegraphics {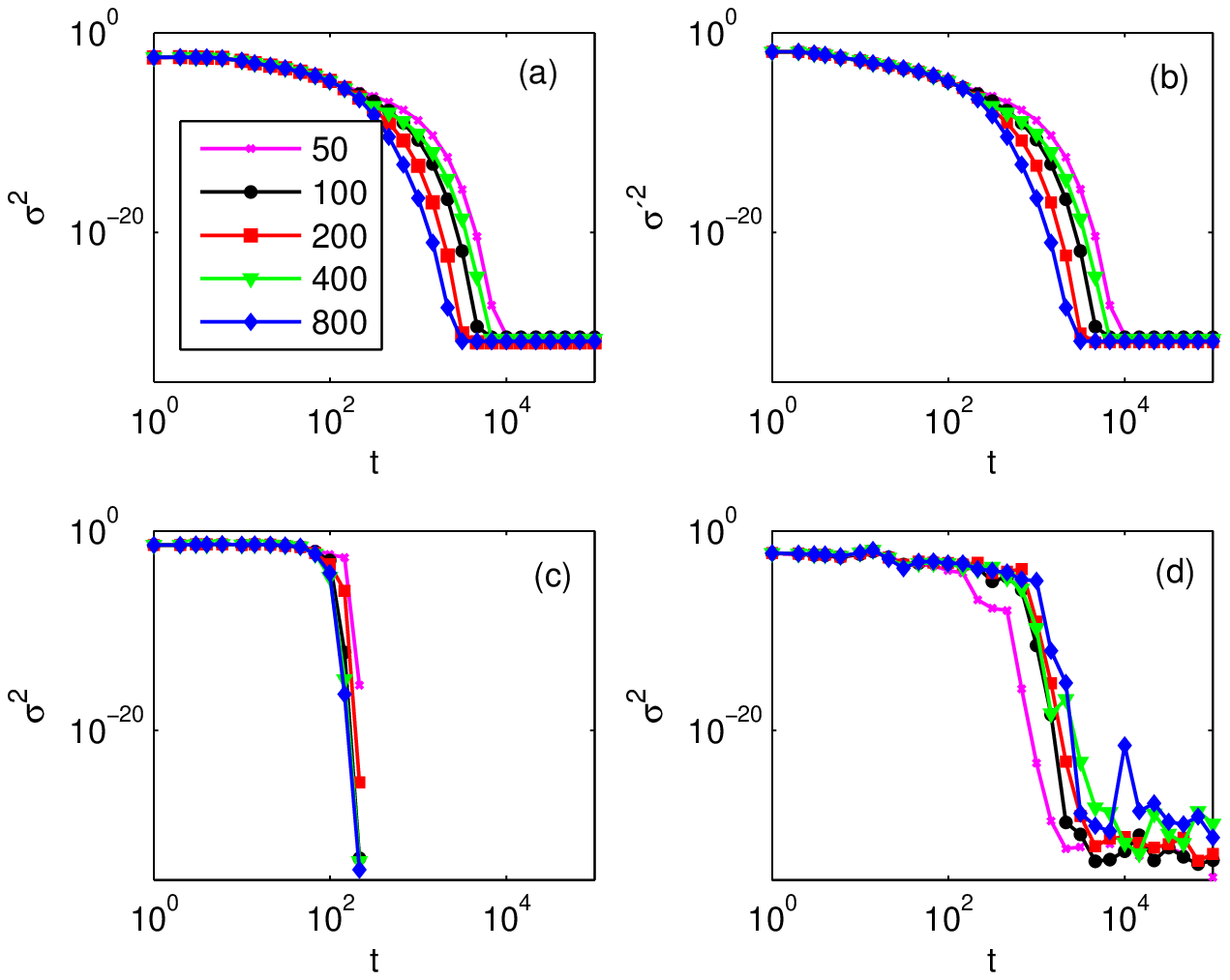}}
\caption{(a), (b) Time-evolution $\sigma^2$ and $\sigma'^2$, during the onset of steady-state synchronization ($\epsilon=0.69$, $c^*=0.57$). (c) Time-evolution of $\sigma^2$ during the onset of time-dependent synchronization for homogeneous delays ($\epsilon=0.45$, $c^*=0$). $\sigma'^2$ remains finite and is not shown. (d) Time-evolution of  $\sigma^2$ during the onset of time-dependent synchronization, in the window for weak coupling existing only for homogeneous and odd delays ($\epsilon=0.18$, $c^*=0$). $\sigma'^2$ remains finite and is not shown. $\sigma^2$ and $\sigma'^2$ were computed for the various network sizes indicated in panel (a).}
\label{fig:trans}
\end{center}
\end{figure}

In Fig.~\ref{fig:trans} we consider finite-size and time-dependent effects during the onset of steady-state synchronization, Figs. ~\ref{fig:trans}(a),(b) and of time-dependent synchronization, Figs. ~\ref{fig:trans}(c),(d). We plot the time-evolution of the instantaneous values of $\sigma^2$ and $\sigma'^2$ [i.e., $\sigma^2$ and $\sigma'^2$ are computed as in Eqs. (\ref{sp1})--(\ref{sp2}) but without time-averaging] for various network sizes $N$, while the average number of neighbors per node is kept constant. Approaching the steady-state synchronization, there is a gradual decrease of the quantifiers, and initially their time-evolution is independent of the network size. In contrast, the approach to time-dependent synchronization, Figs. ~\ref{fig:trans}(c),(d) occurs abruptly, at a time that is nearly independent of the network size.

\begin{figure}[tbp]
\begin{center}
\resizebox{1.0\columnwidth}{!}{\includegraphics {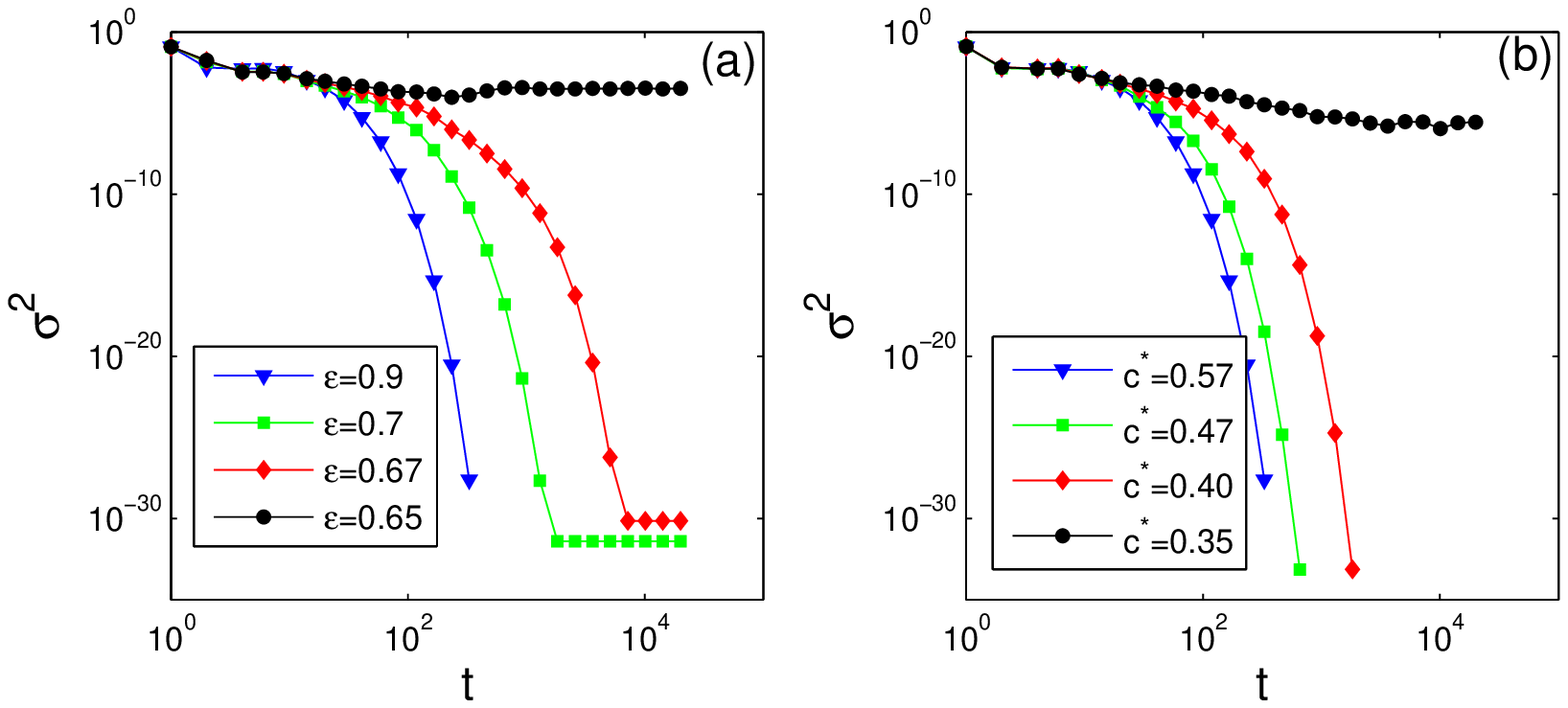}}
\caption{Time-evolution of $\sigma^2$ during the transition to steady-state for (a) various values of $\epsilon$, $c^*=0.57$ (b) for $\epsilon=0.9$ and various values of $c^*$. $\sigma'^2$ exhibits similar behavior (not shown).}
\label{fig:crit_slow_down}
\end{center}
\end{figure}

For parameters close to ``steady-state" synchronization critical slowing down occurs during the approach to the homogeneous steady state, as can be seen in Fig. \ref{fig:crit_slow_down}, where we display the time-variation of $\sigma^2$ for various values of $\epsilon$ and $c^*$.

We have checked the robustness of the above observations by considering delays that are exponentially distributed, and very similar results were found: the formation of two clusters before the onset of steady-state synchronization for increasing $\epsilon$, while there is a single cluster for increasing $c^*$. The small synchronization region that occurs for weak coupling strength is also robust to exponentially distributed delays, as long as the width of the distribution is not too wide. The difference with Gaussian delays is that, with exponentially distributed delays, for strong coupling ($\epsilon\approx 1$), steady state synchronization is lost ($\sigma'^2$ is small and positive) but the network remains synchronized, as $\sigma^2=0$ and the transition probabilities in the nodes are all equal.

\section{Conclusions}
To summarize, we have studied the transition to synchronization in a network of delay-coupled logistic maps. When the coupling delays are homogeneous throughout the network, the network synchronizes to a time-dependent state; when the delays are sufficiently heterogeneous, the synchronization occurs in a steady-state. We employed global and local measures to characterize the synchronization transitions. The global measures are the standard deviation of the distance to the synchronized state, as well as the standard deviation of the transition probabilities in the nodes. The transition probabilities were computed using symbolic analysis and ordinal patterns. We have found that, as the coupling strength increases or as the width of the delay distribution grows, there is a gradual approach to the synchronized state, as seen with the global indicators. An inspection of the local dynamics in the individual nodes, measured by the transition probabilities, reveals that for increasing coupling there is the formation of two clusters before the steady-state synchronization, detected by the fact that the nodes exhibit two qualitatively different transition probabilities. For increasing delay heterogeneity, no cluster formation is seen at the onset of steady-state synchronization, but there is a diversity of values of transition probabilities.

\section{Acknowledgments}
This research was supported in part by the Spanish Ministerio de Educacion y Ciencia through project FIS2009-13360-C03-02, the Agencia de Gestio d'Ajuts Universitaris i de Recerca (AGAUR), Generalitat de Catalunya, through project 2009 SGR 1168, and the ICREA foundation. The authors acknowledge the hospitality of The Max Planck Institute for Physics of Complex Systems, Dresden, where the initial steps of the research were taken.

{}
\end{document}